\documentclass[sigconf]{acmart}

\AtBeginDocument{%
  \providecommand\BibTeX{{%
    \normalfont B\kern-0.5em{\scshape i\kern-0.25em b}\kern-0.8em\TeX}}}

\copyrightyear{2020}
\acmYear{2020}
\setcopyright{acmlicensed}
\acmConference[KDD '20]{The 26th ACM SIGKDD Conference on Knowledge Discovery and Data Mining}{August 22--27, 2020}{San Diego, CA, USA}
\acmBooktitle{The 26th ACM SIGKDD Conference on Knowledge Discovery and Data Mining (KDD '20), August 22--27, 2020, San Diego, CA, USA}

\usepackage{algorithm}
\usepackage[noend]{algpseudocode}
\usepackage{subcaption}
\usepackage{graphicx}
\usepackage{tabulary} 
\usepackage{colortbl,booktabs}
\usepackage{mathrsfs}
\usepackage{threeparttable}
\usepackage{multirow}
\usepackage{booktabs}
\usepackage{bm}

\newcommand{\tabincell}[2]{\begin{tabular}{@{}#1@{}}#2\end{tabular}}

\DeclareMathOperator*{\A}{\mathcal{A}}
\DeclareMathOperator*{\Beh}{\mathcal{B}}
\DeclareMathOperator*{\M}{\mathcal{M}}

\DeclareMathOperator*{\D}{\mathcal{D}}
\DeclareMathOperator*{\Loss}{\mathcal{L}}
\DeclareMathOperator*{\RFunc}{\mathcal{R}}
\DeclareMathOperator*{\vo}{\textbf{o}}

\hypersetup{draft}
\begin{document}

\title[MBCAL: Sample Efficient and Variance Reduced Reinforcement Learning for Recommender Systems]{MBCAL: Sample Efficient and Variance Reduced Reinforcement Learning for Recommender Systems}

\begin{abstract}
In recommender systems such as news feed stream, it is essential to optimize the long-term utilities in the continuous user-system interaction processes. Previous works have proved the capability of reinforcement learning in this problem. However, there are many practical challenges to implement deep reinforcement learning in online systems, including low sample efficiency, uncontrollable risks, and excessive variances. To address these issues, we propose a novel reinforcement learning method, namely model-based counterfactual advantage learning (MBCAL). The proposed method takes advantage of the characteristics of recommender systems and draws ideas from the model-based reinforcement learning method for higher sample efficiency. It has two components: an environment model that predicts the instant user behavior one-by-one in an auto-regressive form, and a future advantage model that predicts the future utility. To alleviate the impact of excessive variance when learning the future advantage model, we employ counterfactual comparisons derived from the environment model. In consequence, the proposed method possesses high sample efficiency and significantly lower variance; Also, it is able to use existing user logs to avoid the risks of starting from scratch. In contrast to its capability, its implementation cost is relatively low, which fits well with practical systems. Theoretical analysis and elaborate experiments are presented. Results show that the proposed method transcends the other supervised learning and RL-based methods in both sample efficiency and asymptotic performances.
\end{abstract}

%\author{Anonymous Authors}
\author{Fan Wang}
\authornote{The authors contributed equally to this research.}
\authornote{Corresponding Authors.}
\email{wang.fan@baidu.com}
\affiliation{\institution{Baidu Inc.}}

\author{Xiaomin Fang}\authornotemark[1]\email{fangxiaomin01@baidu.com}
\affiliation{\institution{Baidu Inc.}}

\author{Lihang Liu}\authornotemark[1]\email{liulihang@baidu.com}
\affiliation{\institution{Baidu Inc.}}

\author{Hao Tian}\authornotemark[2]
\email{tianhao@baidu.com}
\affiliation{
\institution{Baidu Research}}

\author{Zhiming Peng}
\email{pengzhiming01@baidu.com}
\affiliation{
\institution{Baidu Inc.}}

%%\renewcommand{\shortauthors}{Wang, Fang and Liu}






%\setcounter{secnumdepth}{1} %May be changed to 1 or 2 if section numbers are desired.

% The file aaai20.sty is the style file for AAAI Press 
%\setlength\titlebox{2.5in} % If your paper contains an overfull \vbox too high warning at the beginning of the document, use this
% command to correct it. You may not alter the value below 2.5 in
\maketitle

\section{Introduction}
A recommender system (RS) provides users with personalized contents, which significantly improves the efficiency of information acquisition. Nowadays, a typical RS such as news feed stream needs to address multiple steps of user-system interactions in a session. The recommended content in historical interactions can affect the subsequent user behaviors. For instance, exploration of new topics may stimulate the user's interest in related topics; Repeating overlapped contents can make the user lose his/her interest quickly. 
Traditional recommender systems employ collaborative filtering \cite{schafer2007collaborative, koren2009matrix}, or neural networks \cite{cheng2016wide, hidasi2017recurrent, he2017neural} to estimate users' instant behaviors, e.g., instant clicks. However, merely focusing on users' instant behaviors causes many problems, such as overcrowded recommendations, which damage users' experience in the long run.
%Recent works also model user's sequential behaviors using recurrent neural structures \cite{Hidasi2015Session, hidasi2017recurrent, li2017neural}.
Recently there is increased attention on applying deep reinforcement learning (Deep RL) \cite{mnih2015human,lillicrap2015continuous} to recommender systems, which models the user-system interactions as Markov Decision Processes (MDP). A large number of the studies in this area lies in model-free reinforcement learning (MFRL) methods such as Policy Gradient \cite{chen2019top}, DDPG \cite{dulac2015deep}, DRR \cite{liu2018deep}, DeepPage \cite{zhao2018deep} and DQN based recommender systems \cite{zheng2018drn, zhao2018recommendations}. However, many challenges remain in this area. One of them is the over-consumption of data in the training process, which is also referred to as low sample efficiency. Another challenge of MFRL is the practical risks in implementation. On the one hand, on-policy RL can hardly utilize off-policy user logs for training, which raises challenges in online infrastructures and performances at the early stage. On the other hand, off-policy RL suffers from the risk of falling into \emph{Deadly Triad} \cite{Hassselt2018Deep}. It refers to the case of non-convergence when combining off-policy RL with function approximation (such as neural networks) and offline training.

As an alternative choice of MFRL, model-based RL (MBRL) possesses higher sample efficiency and lower practical risks. In MBRL, an environment model (also known as world model) is employed to predict the instant feedbacks and state transitions, and a planning module is implemented to search for an optimal trajectory \cite{Peter2011PILCO}. However, MBRL needs much computation when coming to inference. To make things worse, planning is completely infeasible in a multi-stage retrieval framework, which is widely used in modern recommender systems. In such systems, an earlier stage generates the candidate set of items for the next one; thus, the candidates can not be predetermined. To avoid those issues, authors have proposed to apply Dyna algorithms in recommender systems \cite{liu2019personalized, zoupseudo}. The Dyna algorithm \cite{Sutton1991Dyna} accelerate the convergence through generating virtual interaction by taking advantage of the environment model. However, as a cost of faster convergence, Dyna suffers from losses in asymptotic performance, due to the accumulation of error from the virtual interactions.

\begin{figure*}[t!]
\centering
    \begin{subfigure}{0.66\columnwidth}
        \includegraphics[width=1.0\columnwidth]{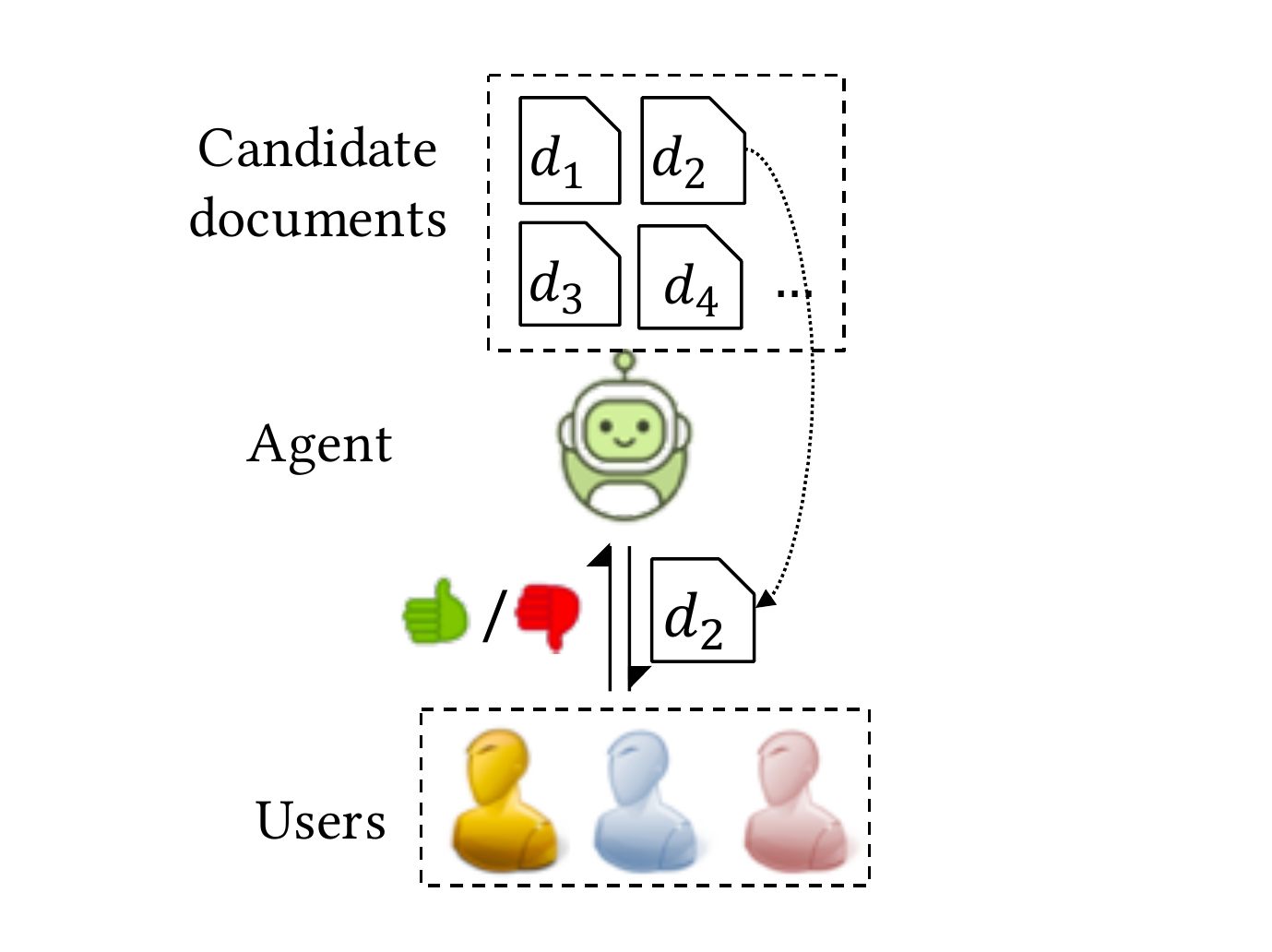}
        %\subcaption{The user-agent interaction}
        \subcaption{}
        \label{fig:Intro_1}
    \end{subfigure}
    \begin{subfigure}{0.66\columnwidth}
        \includegraphics[width=1.0\columnwidth]{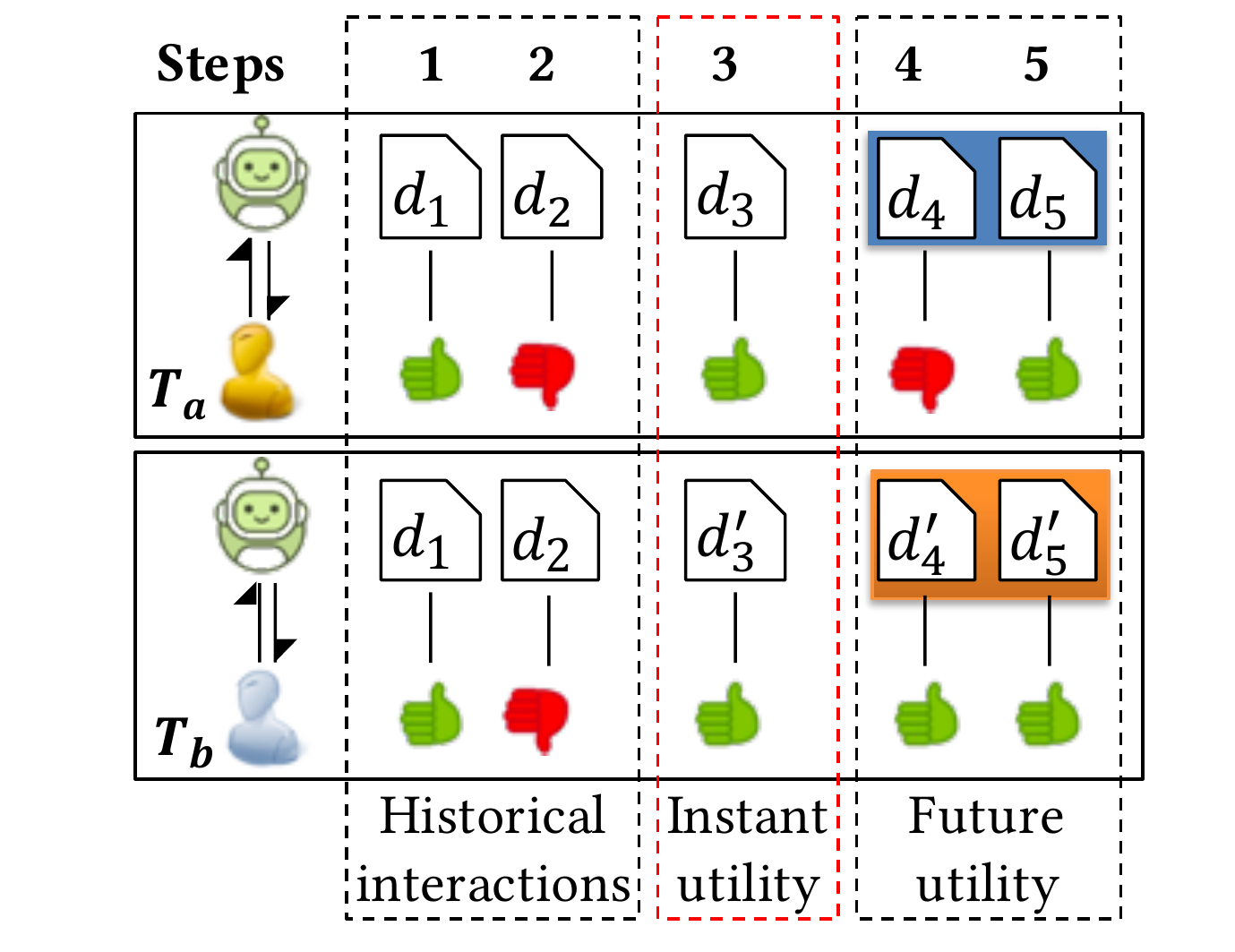}
        %\subcaption{Examples of the interactive trajectories}
        \subcaption{}
        \label{fig:Intro_2}
    \end{subfigure}
    \begin{subfigure}{0.66\columnwidth}
        \includegraphics[width=1.0\columnwidth]{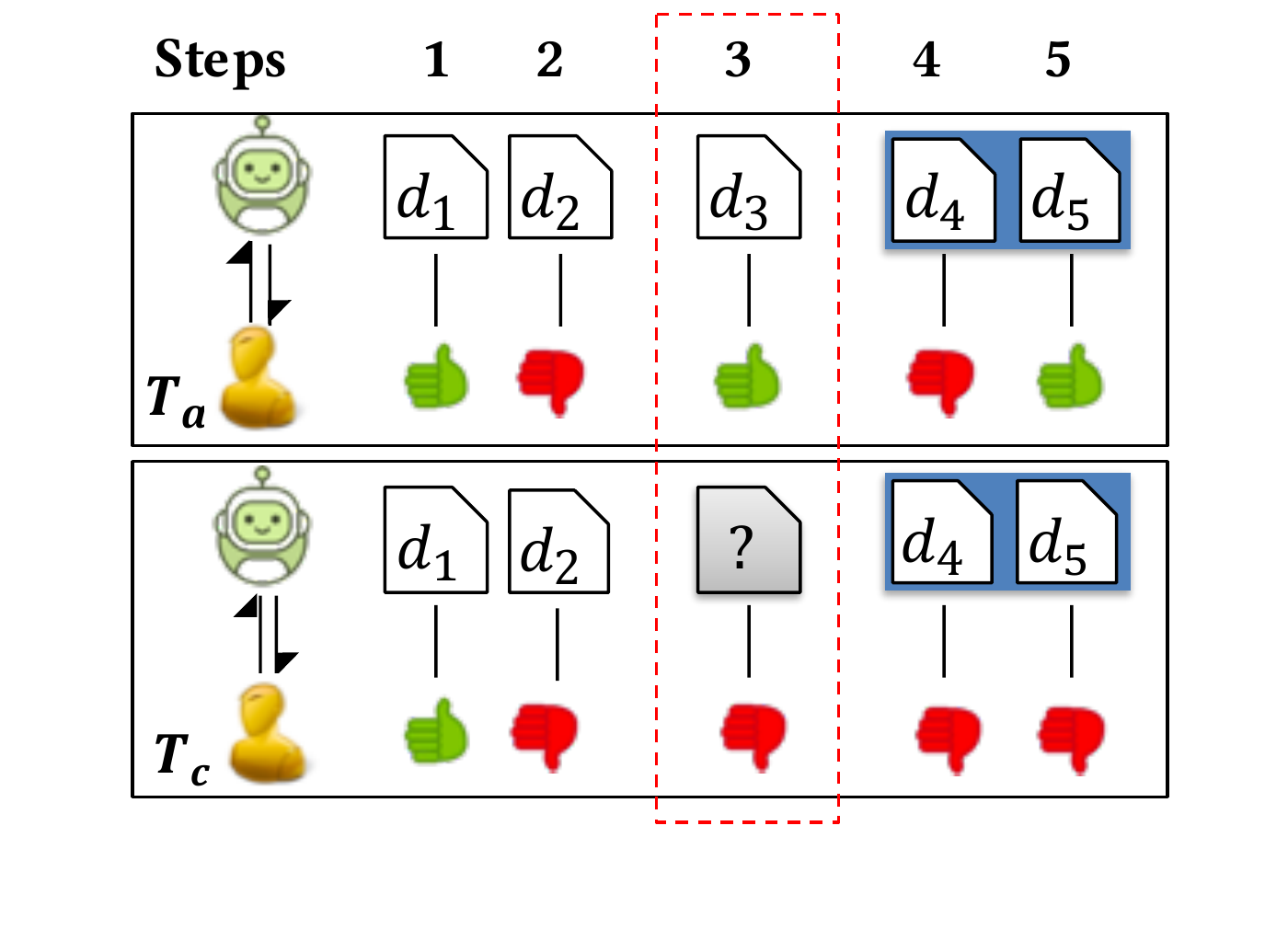}
        %\subcaption{Examples of the interactive trajectories}
        \subcaption{}
        \label{fig:Intro_3}
    \end{subfigure}
\caption{(\subref{fig:Intro_1}) Illustration of the user-system interaction; (\subref{fig:Intro_2}) Two examples of trajectories of interaction, $T_a$ and $T_b$, generated by different users; (\subref{fig:Intro_3}) The original trajectory $T_a$ and couterfactual comparison $T_c$ from the same user}.
\label{fig:intro}
\vspace{-0.15in}
\end{figure*}

Excessive variance of gradients in optimization is an important challenge of deploying RL as well. The variance may originate from stochastic transition, noisy rewards, and stochastic policy. Longer horizons are found to exacerbate the variance. Excessive variance significantly slows down the convergence and introduces instabilities. Previous work \cite{schulman2016high, Ziyu2016Dueling} has shed light on this issue by showing that using advantage function instead of value function can reduce the variance and thus improve the performance. However, those proposals aim at MFRL, and variance reduction in MBRL has not been studied yet.

Specifically, in recommender systems, the variance may come from the following aspects. First, there are extensive noises in the observed user feedbacks. For example, some users are more prone to give positive/negative feedbacks than the others. Even for a single user, he/she may behave differently at different times of a day (e.g., before sleeping vs. at work). Second, for stochastic policy, re-sampling the trajectory starting from any state can lead to variant long-term returns. Although the influence of variances can be alleviated by inducing from a sufficiently large amount of data, the variances still have negative impacts due to the sparsity of the data for specific user and item.

To clearly explain the influence of variances, we show a demonstration of the interaction process in Figure~\ref{fig:intro}(\subref{fig:Intro_1}). For each step, the agent selects an item for display, and the user returns his/her feedbacks. An \emph{observed trajectory} includes multiple steps of interactions, as $T_a$ and $T_b$ shown in Figure~\ref{fig:intro}(\subref{fig:Intro_2}). In our settings, the candidate user behaviors include "Like (Thumbs Up)" and "Unlike (Thumbs Down)", and the utility is the number of "Likes" in the trajectory. Here, we consider the document $d_3$ in the 3rd step in trajectory $T_a$, which yields 2 "Likes" considering the instant and future utilities. It is hard to judge whether $d_3$ is a superior action here due to the lack of comparison. To give a better evaluation, we can find a comparison such as Trajectory~$T_b$ in Figure~\ref{fig:intro}(\subref{fig:Intro_2}). $T_b$ possesses the same historical interactions as $T_a$, which implies that the user share many interests with that of $T_a$. The trajectory starting from $d'_3$ in $T_b$ yields 3 following "Likes", by which we may judge that $d'_3$ is better than $d_3$ . However, it is a quite arbitrary conclusion, because the difference in future utility can possibly be attributed to the user biases, or the quality of follow-up items ($d_4, d_5$ vs. $d'_4, d'_5$).
%First, though the last three steps in B1 get three "Unlike", this can be caused by the quality of the following items ($d_9,d_4,d_{10}$) instead of $d_2$. Second, as A1 and B1 are generated separately by different users, it is possible that the two users tend to behave differently in the latter part of the session. Even if those trajectories are generated by the same user, there may still be other contextual factors to be blamed, e.g., A1 is generated at night and B1 is generated during the work hour. 

Aiming at further reducing the variances, a key idea of our work is to compare $T_a$ with another trajectory, $T_c$ in Figure~\ref{fig:Intro_3}. $T_c$ shares all the contexts with $T_a$, including the user, historical interactions, and follow-up items ($d_4, d_5$), except for replacing the current document $d_3$ with some other documents. By comparing $T_a$ with $T_c$, we can come to a more solid conclusion on the advantage of taking $d_3$. Unfortunately, it is impossible to find records such as $T_c$ from user logs, as a user can not possibly go through the same trajectory twice. However, by taking advantage of the environment model, we can do simulations into the future (often referred to as simulated rollout), and we can indeed generate trajectories like $T_c$.

Following the idea mentioned above, we propose a novel MBRL solution toward RS, namely the \textbf{Model-based Counterfactual Advantage Learning} (MBCAL). First, the overall utility is decomposed into the instant utility (the rewards acquired in the current step) and the future utility (the rewards acquired in the future). The instant utility is naturally predicted with the environment model, and the future utility is approximated through simulated rollout. Second, to further reduce the variance in the future utilities, we try to do two comparative simulated rollouts. Before doing so, we introduce the masking item to the environment model, which allows us to generate simulated rollouts by masking the document in the step that we are interested in (the trajectory $T_c$). We then calculate the counterfactual future advantage (CFA) as the difference of the future utility with and without masking. At last, we introduce the future advantage model to approximate the CFA.
%The priority to select an action in inference is the combination of the environment model and the future advantage model.

We conduct simulative experiments by utilizing three real-world datasets. The methods for comparison include supervised learning, MFRL and MBRL. We also put our attention on Batch-RL and Growing Batch-RL settings\cite{lange2012batch}, which is more compatible with the practical infrastructures. Extensive
results of experiments show the superiority of the proposed method.

\section{Preliminaries}

\subsection{Problem Settings and Notations}\label{sec:notation}

We formalize the recommendation as Markov Decision Processes (MDP), denoted with the symbols $(\mathcal{S}, \mathcal{A}, \mathcal{R}, \mu)$. The details are explained as follows:
\begin{itemize}
\item \emph{State} $s_t \in \mathcal{S}$ represents the unknown user interests and contextual information at step $t$. As it is impossible to know the user interest exactly, the observed user interaction history is frequently used as the state descriptor \cite{zhao2019toward, liu2018deep}, i.e.,
\begin{equation}
s_t = (o_1, o_2, ..., o_{t-1})
\label{eq:POMDP_Approximation}
\end{equation}
where we use $o_t$ to represent the pair of exposed item $a^{\vo}_t$ and corresponding feedback $b^{\vo}_t$, i.e., $o_t=(a^{\vo}_t, b^{\vo}_t)$. For simple we also use $\vo\nolimits_{[1:t-1]}$ to represent the trajectory $o_1, o_2, ..., o_{t-1}$.
\item \emph{Action} $a_t \in \A_t$ denotes the selected item by the agent, with $\A_t$ being the candidate set that is passed from the earlier stage of the recommender system.
\item \emph{Reward} $r_t \in \mathcal{R}$ is the utility that we want to maximize. Usually, it depends on the observed user behaviors.
%\item \emph{Observation} $o_t \in \mathcal{O}$ in recommender system includes the actions that are exposed to the user, and the corresponding behaviors from the user. 
\item \emph{Conditional Transition Probabilities} $\mu(s'|s,a)$ denotes the transition probability to state $s'$ given state and action pair $(s,a)$, which is hidden in most cases.
%\item \emph{Conditional Observation Probability} $\Omega(o|s)$ denotes the probability of observe $o$ at state $s$, which is also hidden.
\end{itemize}

\begin{table}
  \caption{Notations.}
  \begin{tabular}{l | l}
    \hline
    \textbf{Notations} & \textbf{Descriptions} \\
    \hline
    $\Beh$ & \tabincell{l}{Categories of user behaviors, e.g., \\ $\Beh$=("Skip", "Click", "Click and Thumbs up")}\\
    \hline
    $\RFunc$ & \tabincell{l}{Rewards corresponding to the \\behaviors in $\Beh$, e.g., $\RFunc=(0, 1, 5)$}\\
    \hline
    $r$ & $r \in \RFunc$, reward / utility \\
    \hline
    $s$ & state \\
    \hline
    $a$ & action \\
    \hline
    $\pi(a|s)$ & policy for the recommendation \\
    \hline
    $r(s,a)$ & reward function \\
    \hline
    $\mu(s'|s,a)$ & transition probabilities \\
    \hline
    $Q^{\pi}(s,a)$ & state-action value function of $\pi$ \\
    \hline
    $V^{\pi}(s)$ & state value function of $\pi$ \\
    \hline
    $\vo$ & \tabincell{l}{observed interactive trajectory \\ $\vo = (a^{\vo}_1, b^{\vo}_1, a^{\vo}_2, b^{\vo}_2$, ...)}\\
    \hline
    $\vo_{[t_1:t_2]}$ & sub-trajectories from step $t_1$ to $t_2$ \\
    \hline
    $a^{\vo}_t$ & action taken at $t$ in trajectory $\vo$ \\
    \hline
    $b^{\vo}_t$ & $b^{\vo}_t \in \Beh$, user behavior observed at $t$ in $\vo$ \\
    %\hline
    %$\M(\vo, y, a) $ & \tabincell{l}{trajectory acquired by replacing the action\\ at steps $y$ in $\vo$ with $a$} \\
    \hline
    $\D^{\pi}$ & collection of trajectories with policy $\pi$ \\
    \hline
  \end{tabular}
  \label{tab:notation}
\end{table}

Additionally, we denote the user feedback (behavior) at step $t$ with $b_t \in \Beh$, where $\Beh = \{\Beh_1, \Beh_2, ..., \Beh_n\}$ is the set of candidate user behaviors (e.g., "Click", "Skip", "Click and Thumbs Up"). 
Also, notice that we use the superscript $\vo$ to denote that an action $a_t$ and user feedback $b_t$ belongs to a specific trajectory, given $a^{\vo}_t$ and $b^{\vo}_t$. Without losing generality, we focus on maximizing the overall utility within a fixed $T$ steps of interactions. In addition to the symbols mentioned above, we also adopt other commonly used notations in RL, shown in Table~\ref{tab:notation}.

\subsection{RL-based Recommender Systems}
\label{sec:rl_background}
%As addressing POMDP may be complicated, most previous studies of RL-based recommender systems directly treat the observed trajectory as the state 
%In such a way, POMDP degenerates to MDP. In this paper, we continue to use this proposition.
%However, unlike the previous work that merely uses Equation~\eqref{eq:POMDP_Approximation} as a feature descriptor, using a sequence of observations to represent the state is essential to MBCAL. We explain the details in Section~\ref{sec:methodology}. 
Without confusion, we replace state $s_t$ with the trajectory $\vo_{[1:t-1]}$, and denote the value and policy function as $Q^{\pi}(\vo_{[1:t-1]}, a)$ and $\pi(a|\vo_{[1:t-1]})$, respectively. It is worth to notice that other user-side features (user-ids, user properties, context features such as time of the day) can be involved in the state representation without any difficulty. However, for simplicity, we omit the notation of those features.

Typical MFRL methods try to learn the function approximator denoted as $Q_{\theta}$, with $\theta$ being trainable parameters, which is optimized by minimizing the loss function:
\begin{equation}
\Loss_{\text{MFRL}}(\theta) = \frac{1}{|\D|}\sum_{\vo \in \D}\frac{1}{T}\sum_{t=1}^{T}[\hat{Q}_{\text{target}} - Q_{\theta}({\vo}_{[1:t-1]}, a^{\vo}_t)]^2.
\label{eq:ValueIteration}
\end{equation}
$\hat{Q}_{\text{target}}$ here represents the value backup (backup: updating the value of current states with that of future states \cite{sutton1998introduction}), which is calculated differently in different algorithms. E.g., Deep Q Network (DQN) \cite{mnih2015human} uses the temporal difference errors, where $\hat{Q}_{\text{target}}=$ $r_t + \gamma \max_{a'}Q_{\theta'}(\vo_{[1:t]},a')$. $\theta'$ here is the set of parameters which is periodically copied from $\theta$ during the optimization process.

\subsection{Growing Batch-RL}
%We suppose that each user interacts with the environment for a horizon of $T$ \emph{step}s, namely an \emph{episode}. In each step, the recommendation system selects an item from the candidates and exposes the item to the user, after which the system observed the user behavior toward the exposed item. A predefined reward function decides the reward given the specific user behavior. This process continues until $t=T$. The objective is to maximize the overall rewards of an episode.
% We focus on optimizing the recommendation strategy in this paper. 
%\begin{figure}
%\centering
%    \includegraphics[width=0.60\columnwidth]{images/BatchRL.png}
%\caption{A Sketch of Growing Batch Reinforcement Learning}
%\label{fig:BatchRL}
%\end{figure}

As online update is typically not achievable in realistic recommender systems, we are more interested in Batch-RL and Growing Batch-RL \cite{lange2012batch} settings. Batch-RL refers to the policy learning with static user logs. It can usually be used to evaluate the capability of utilizing the offline data.
Growing Batch-RL lies somewhere in between the Batch-RL and online learning. The data can be re-collected in a batch manner and used for policy update iteratively. Many recently proposed RL-based systems use a framework that is close to Growing Batch RL settings \cite{zheng2018drn, zhao2018deep,chen2019top}. More specifically, Growing Batch RL includes two periodic interleaving stages:

\textbf{Data Collection}. The agent uses the policy $\pi_k$ to interact with the environment (the users). The policy is kept static during this process. The collected interactive trajectories are denoted as $\D^{\pi_k}$.

\textbf{Policy Update}. The interactive trajectories $\D^{\pi_k}$ are used for training, the agent update the policy from $\pi_k$ to $\pi_{k+1}$.

The essential problems of Batch-RL and Growing Batch-RL are to guarantee the \textbf{Policy Improvement}, i.e., how much the performance of $\pi_{k+1}$ is improved compared with $\pi_k$.

\section{Methodology} 
\label{sec:methodology}

\begin{figure}
\centering
    \includegraphics[width=0.95\columnwidth]{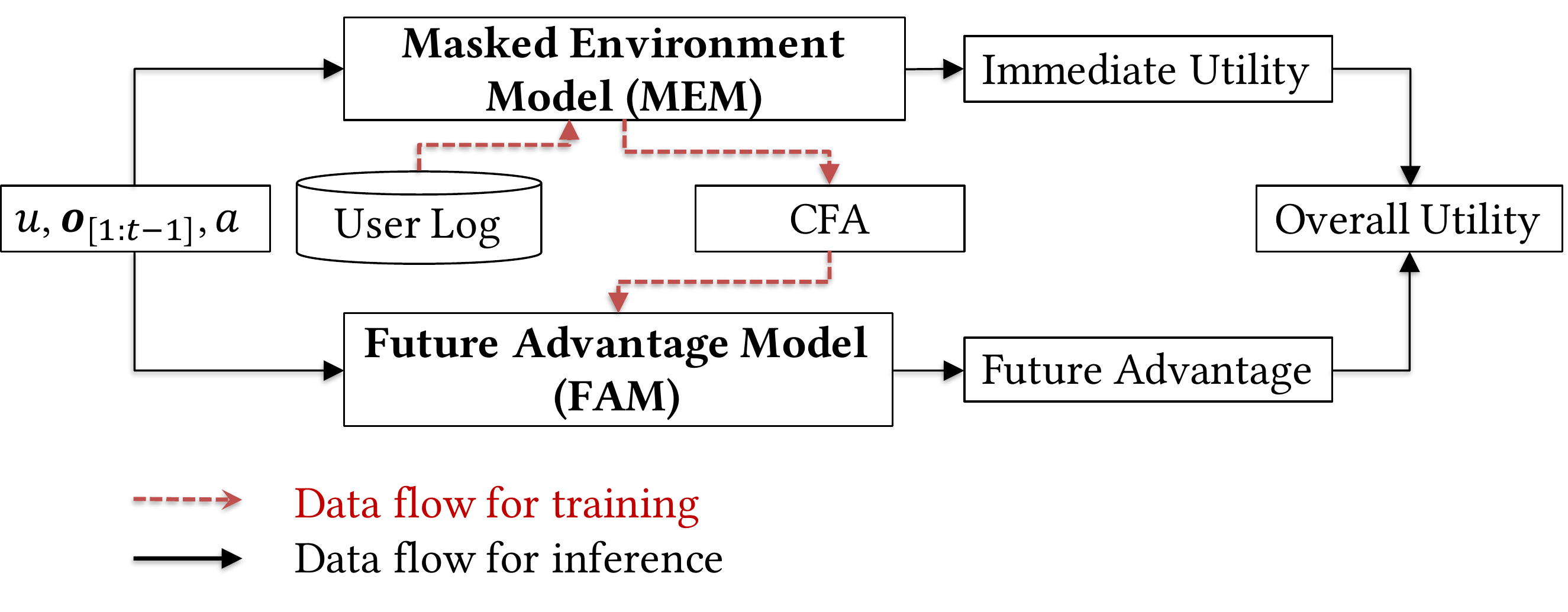}
\caption{Overview of MBCAL}
\label{fig:MBCAL}
\end{figure}

%In the following part, we will first explain the MBCAL procedure in detail. Then we will give a theoretical validation of the proposed method, with also remarks on its strengths and limitations.
The fundamental idea of MBCAL is explained with Figure~\ref{fig:MBCAL}. We use two models to approximate the instant user behavior and the future advantage separately: the Masked Environment Model (MEM) and the Future Advantage Model (FAM). For the training, we start with optimizing the environment model to predict user behaviors, where we also introduce the masking item into the model. With the MEM, we can calculate the Counterfactual Future Advantage (CFA) by comparing the future utility of masking the action or not. CFA further serves as the label of FAM. For the inference, we use the combination of the two models to pick actions.

In this section, we first formalize the environment model, and then we describe MEM, FAM, and the learning process overall. They are followed by the theoretical analysis of the proposed method.

\subsection{Environment Modeling}
\label{sec:model_based_rl}

As the most common setting, an environment model predicts the transition and the reward separately. Here, we use $\hat{\mu}(s_t,a_t,s_{t+1})$ to approximate $\mu(s_{t+1} | s_t,a_t)$ and $\hat{r}(s_t, a_t)$ to approximate the reward $r(s_t, a_t)$. Specifically, to derive the formulation of environment model in RS, we use Equation~\eqref{eq:POMDP_Approximation}, and $\mu(s_{t+1}|s_t, a_t)$ can be rewritten by Equation~\eqref{eq:transition}.
\begin{align}
\mu(s_{t+1}|s_t, a_t) &=\mu({\vo}_{[1:t]}|{\vo}_{[1:t-1]}, a_t^{\vo}) \nonumber\\
&=\mu({\vo}_{[1:t-1]}, a_t^{\vo}, b_t^{\vo}| {\vo}_{[1:t-1]}, a_t^{\vo}) \nonumber\\
&=\mu(b_t^{\vo}|{\vo}_{[1:t-1]}, a_t^{\vo})
\label{eq:transition}
\end{align}
In other words, the prediction of transition degenerates into the prediction of instant user behaviors. Notice that reward is dependent on the user behavior only as well, thus it is possible to use only one model in place of $\hat{\mu}$ and $\hat{r}$. We introduce $f_{\phi}(\vo\nolimits_{[1:t-1]} , a^{\vo}_t, b_t^{\vo})$ with trainable parameters $\phi$ to approximate $\mu(b_t^{\vo}|{\vo}_{[1:t-1]}, a_t^{\vo})$, which predicts the probability of the next user behavior being $b_t^{\vo}$. The transition and the reward are then naturally approximated with
\begin{align}
&\hat{\mu}(s_t, a_t, s_{t+1}) = f_{\phi}(\vo\nolimits_{[1:t-1]} , a^{\vo}_t, b^{\vo}_t).\\
&\hat{r}(s_t, a_t) = \sum_n {\RFunc}_{n} \cdot f_{\phi}(\vo\nolimits_{[1:t-1]} , a^{\vo}_t, \Beh\nolimits_n),
\label{eq:EnvironmentModel}
\end{align}

%Here we use the notation $r_{\phi}()$, as it is completely dependent on $f_{\phi}$, and thus they share the parameter $\phi$. 

\subsection{Masked Environment Model}

To eliminate the intractable noises hidden in the feedbacks, we introduce a masking item into the model. The motivation of this is to try to find a counterfactual comparison to the current trajectory, which answers the question: "If this action was not taken, what would the future behavior be like?" To do so, we introduce a virtual item $a_M$, which is represented by a trainable embedding vector. For convenience, given an observation trajectory $\vo$, we denote the trajectory where the actions at positions $y=\{t_1,t_2,...\}$ are replaced by $a_M$ as $\M(\vo, y, a_M)$. For example, suppose $y = \{ 2, 3\}$, it gives a masked trajectory of
\begin{equation*}
    \begin{split}
        & \M(\vo,\{2,3\}, a_M) = (a^{\vo}_1, b^{\vo}_1, a_M, b^{\vo}_2, a_M, b^{\vo}_3, a^{\vo}_4, b^{\vo}_4, ...),
    \end{split}
\end{equation*}
where actions $a_2^{\vo}$ and $a_3^{\vo}$ in $\vo$ are replaced by $a_M$.

The training is straightforward. We sample random positions $y^{\vo}$ for each trajectory $\vo$, such that each position has uniform probability of $p_{\text{mask}}$ to be replaced. We want the MEM to recover the user behavior as close as possible in case some items are masked. With the collected masked trajectories $\D_M = \{\M(\vo,y^{\vo},a_M)| \vo \in \D\}$, we maximize the likelihood, or minimize the negative log-likelihood (NLL).
\begin{equation}
    \begin{split}
        & \Loss\nolimits_{\text{MEM}}(\phi) = - \frac{1}{|\D_M|}\sum_{\vo \in \D_M}\frac{1}{T}\sum_{t=1}^T[\log{f_{\phi}(\vo\nolimits_{[:t-1]}, a^{\vo}_t, b^{\vo}_t)}].
    \end{split}
    \label{eq:environment_loss}
\end{equation}

To model the sequential observations, the architecture of MEM follows that of session-based recurrent RS(\cite{Hidasi2015Session, hidasi2017recurrent}). We use Gated Neural Network\cite{Graves2013Speech} to encode the trajectory $\vo_{[1:t-1]}$. As we need to encode $\vo_{[1:t-1]}$ and $a^{\vo}_t$ at the same time, we concatenate the input in a staggered way, equivalent to the setting of \cite{santoro2016meta}. For each step $t$, the model takes $b^{\vo}_{t-1}$ and $a^{\vo}_t$ as input and output the probability of the next possible behavior. An additional $b_s$ is introduced as the start of the observed user behavior (see Figure~\ref{fig:Model_Arch}). Concretely, the architecture is formulated as follows.
\begin{align}
    & h_0^{MEM} = \textbf{Emb}(u),\\
    & x_t^{MEM} = \textbf{Concat}(\textbf{Emb}(b_{t-1}), \textbf{Emb}(a_t)), \\
    & h_t^{MEM} = \textbf{GRU}(h_{t-1}^{MEM}, x_t^{MEM}), \\
    & f_{\phi} = \textbf{Softmax}(\textbf{MLP}(h_t^{MEM})).
\end{align}
Here $\textbf{Emb}$ denotes a representation layer; $\textbf{Concat}$ denotes a concat operation and \textbf{MLP} denotes multilayer perceptron; $\textbf{GRU}$ represents a Gated Recurrent Unit.

\subsection{Counterfactual Future Advantage}
With the MEM, we can estimate the difference in future utilities between the original trajectory and the counterfactual comparison, namely the Counterfactual Future Advantage (CFA). Specifically, given the trained MEM $f_{\phi}$, we first define the Simulated Future Reward (SFR, denoted with $\hat{R}^{\text{future}}$) of the observed trajectory $\vo$ at time step $t\in[1,T]$ as
\begin{equation}
    \begin{split}
        \hat{R}^{\text{future}}_{\phi}(\vo, t) = \sum_{\tau=t+1}^{T}\gamma^{\tau-t}r_{\phi}(\vo\nolimits_{[1:\tau-1]}, a^{\vo}_{\tau}).
    \end{split}
    \label{eq:future_reward}
\end{equation}
We then calculate CFA (denoted with $\hat{A}^{\text{future}}$) by subtracting the SFR of counterfactual comparison from the original one, see Equation~\eqref{eq:couterfactual_future_advantage}.
\begin{equation}
    \begin{split}
        & \hat{A}^{\text{future}}_{\phi}(\vo, t) = \hat{R}^{\text{future}}_{\phi}(\vo, t) - \hat{R}^{\text{future}}_{\phi}(\M(\vo, \{t\}, a_M), t).
    \end{split}
    \label{eq:couterfactual_future_advantage}
\end{equation}
% We put some remarks here. $\hat{R}^{\text{future}}_{\phi}(\vo, t)$ is an approximation of future rewards. However, just like Monte Carlo Policy Evaluation, the variance of such evaluation is large due to the noisy stochastic transitions. Baseline term $\hat{R}^{\text{future}}_{\phi}(\vo\nolimits_{\{t\} \rightarrow a_M}, t)$ in Equation~\eqref{eq:couterfactual_future_advantage} is the key to reduce the variance. By subtracting a baseline term, we eliminate the variance brought by uncertain future recommendations, but leave only the influence of the current item on the following items. However, the CFA and CFR is not the actual advantage or value function, but a virtual estimation from the environment model, thus we use the name "Counterfactual".
Finally, we introduce the Future Advantage Model (FAM) denoted with $g_{\eta}(\vo_{[1:t-1]}, a_t)$, with trainable parameters $\eta$. To train FAM, we minimize the mean square error shown in Equation~\eqref{eq:CFAApproximator}.
\begin{align}
        \Loss\nolimits_{\text{FAM}}(\eta) =& - \frac{1}{|\D|}\sum_{(\vo) \in \D}\frac{1}{T}\sum_{t=1}^{T} [g_{\eta}(\vo\nolimits_{[1:t-1]}, a^{\vo}_{t}) \nonumber \\
        &- \hat{A}^{\text{future}}_{\phi}(\vo, t)]^2.
    \label{eq:CFAApproximator}
\end{align}

FAM takes the equivalent input as the MEM. We use the same neural architecture as MEM except for the last layer, but with different parameters. For the last layer FAM predicts a scalar (the advantage) instead of a distribution, shown as follows:
\begin{align}
    & h_0^{FAM} = \textbf{Emb}(u),\\
    & x_t^{FAM} = \textbf{Concat}(\textbf{Emb}(b_{t-1}), \textbf{Emb}(a_t)), \\
    & h_t^{FAM} = \textbf{GRU}(h_{t-1}^{FAM}, x_t^{FAM}), \\
    & g_{\eta} = \textbf{MLP}(h_t^{FAM}).
\label{eq:FAM}
\end{align}

%It is worth to take a closer inspection toward $\Loss_{\text{MFRL}}$ in Equation~\eqref{eq:ValueIteration} and $\Loss_{\text{CFA}}$ in Equation~\eqref{eq:CFAApproximator}. They share the similar formulation except for the targets($\hat{Q}_{\text{target}}$ and $\hat{A}^{\text{future}}_{\phi}$). Both targets try to account for the long term rewards, but $\hat{A}^{\text{future}}_{\phi}$ is remarkable in three aspects: Firstly, $\hat{A}^{\text{future}}_{\phi}$ accounts for the advantage only, while $\hat{Q}_{\text{target}}$ involves the baseline value function. Secondly, $\hat{A}^{\text{future}}_{\phi}$ accounts for the residual of advantage function by taking out the immediate reward of step $t$. This is because the immediate reward has already been captured by the environment model $f_{\phi}$, and there is no need to involve it any more. Thirdly, while $\hat{Q}_{\text{target}}$ uses value backup from real experiences, $\hat{A}^{\text{future}}_{\phi}$ comes from the MEM, which already has lower variances compared with the real interactions. Those features enable $\hat{A}^{\text{future}}_{\phi}$ to capture the long-term rewards with lower noises.

\subsection{Summary of MBCAL}

\begin{figure}[t!]
\centering
\includegraphics[width=\columnwidth]{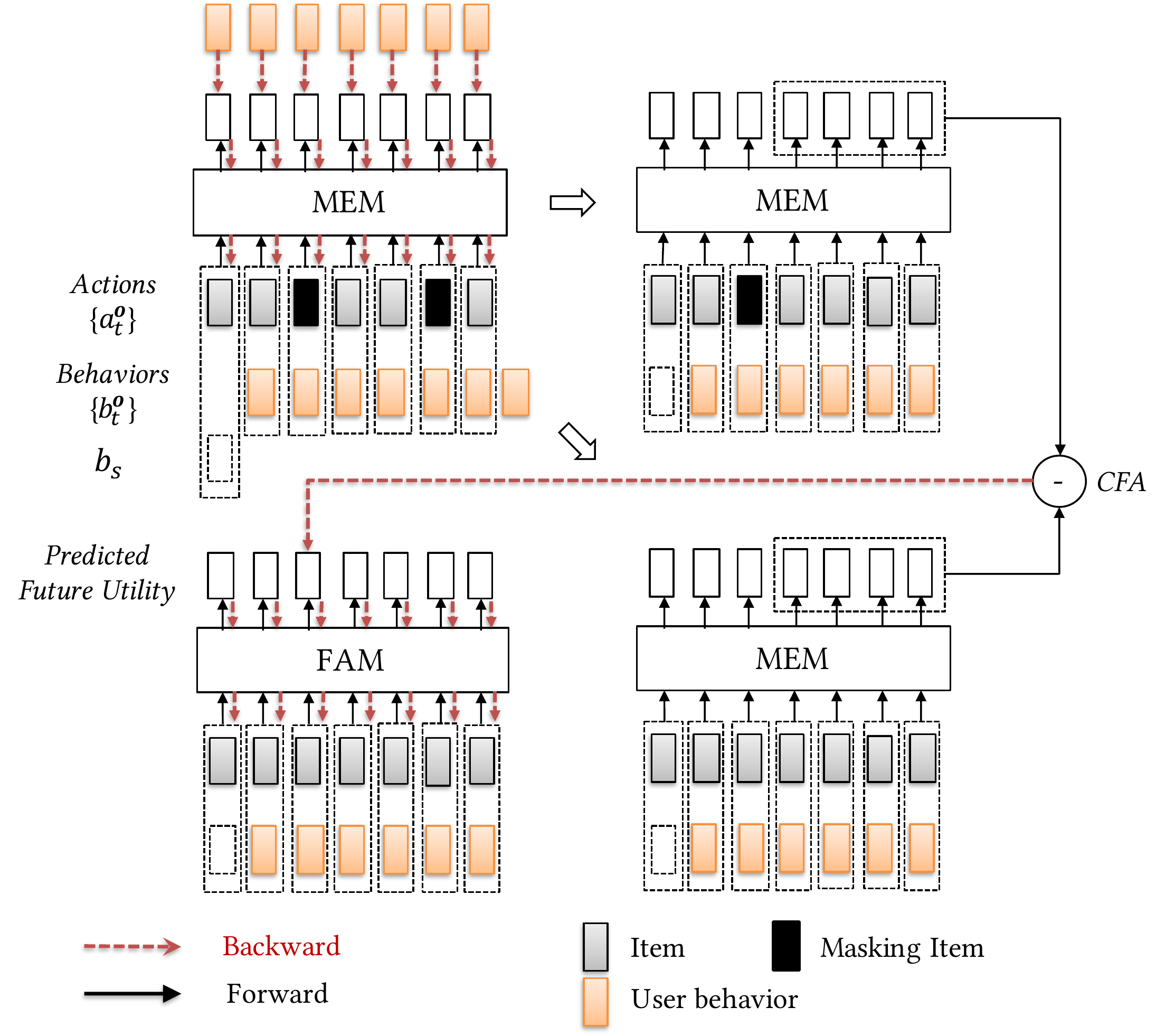}
    %\begin{subfigure}{0.95\columnwidth}
    %    \includegraphics[width=0.9\columnwidth]{images/MEM_Learning.png}
    %    \subcaption{Learning of the masked environment model(MEM).}
    %    \label{fig:MEMLearn}
    %\end{subfigure}
    %\begin{subfigure}{0.95\columnwidth}
    %    \includegraphics[width=0.9\columnwidth]{images/CFA_Learning.png}
    %    \subcaption{Structure of the Counterfactual Future Advantage Approximator(FAM).}
    %    \label{fig:CFALearn}
    %\end{subfigure}
    %\\
    %\begin{subfigure}{1.9\columnwidth}
    %    \includegraphics[width=0.9\columnwidth]{images/CFA_Calculation.png}
    %    \subcaption{Calculating Counterfactual Future Advantage(CFA) based on the trained Masked Environment Model.}
    %    \label{fig:CFACal}
    %\end{subfigure}
\caption{Illustration of the MBCAL architectures.}
\label{fig:Model_Arch}
\end{figure}

An integrate illustration of MBCAL is shown in Figure~\ref{fig:Model_Arch}. For the inference, we select the item(action) based on both the MEM and FMA. Formally, given the user information $u$ and the observation trajectory $\vo_{[1:t-1]}$, we pick the next action according to Equation~\eqref{eq:Recommendation_policy}.
\begin{equation}
    a_t^* = argmax_a [r_{\phi}(\vo\nolimits_{[1:t-1]}, a) + g_{\eta}(\vo\nolimits_{[1:t-1]}, a)].
    \label{eq:Recommendation_policy}
\end{equation}

To avoid the local optimum in policy improvement, we employ $\epsilon$-greedy strategy \cite{sutton2018reinforcement}. With probability $\epsilon$, we select a random action, and with the left probability, we select according to Equation~\eqref{eq:Recommendation_policy}. It is written as
\begin{equation}
    a_t = \left \{
    \begin{aligned}
    &a_t^*, &with \ probability \ 1-\epsilon; \\
    &random \ action \ a \in {\A}_t, &with \ probability \ \epsilon.
    \end{aligned}
    \right.
    \label{eq:Recommendation_policy_explore}
\end{equation}

MBCAL fits well with the Growing Batch-RL settings. The summary of the algorithm is shown in Algorithm~\ref{Algorithm:MBCAL}, with the function \emph{PolicyUpdate} shown in Algorithm~\ref{Algorithm:MBCAE-PL}. Although we use the notation $\pi$ to represent the policy, we do not require any explicit formulation of the policy; The common policy gradient type algorithms require the explicit form of policy, which is hard to acquire in many recommender systems.

\begin{algorithm}
	\caption{Model-Based Counterfactual Advantage Learning(MBCAL)}
	\label{Algorithm:MBCAL}
	\begin{algorithmic}[1]
    \State Initial Policy $\pi_0$.
	\For{k = 0,1,2,... until convergence}
	\State Use policy $\pi_k$ to interact with users for $N$ trajectories, and record the interaction history $\D^{\pi_k}$.
	\State $f_{\phi}, g_{\eta} =\emph{PolicyUpdate}(\D^{\pi_k})$.
	\State Set $\pi_{k+1}$ according to  Equation~\eqref{eq:Recommendation_policy_explore}.
	\EndFor
	\State \textbf{Return} $\pi_{k+1}$ in the last iteration.
	\end{algorithmic}
\end{algorithm}

\begin{algorithm}
	\caption{\emph{PolicyUpdate}($\D$)}
	\label{Algorithm:MBCAE-PL}
	\begin{algorithmic}[1]
    \State \textbf{Input:} Interaction history $\D$.
	\State Randomly masking $\D$ to acquire $\D_M$.
	\State Minimize $\Loss_{\text{MEM}}$(Equation~\eqref{eq:environment_loss}) on $\D_M$ to optimize $f_{\phi}$.
	\State For $\vo \in \D$, calculate $\hat{A}_{\vo, t}$ with $t\in[1,T]$ by Equation~\eqref{eq:future_reward} and Equation~\eqref{eq:couterfactual_future_advantage}
	\State Minimize $\Loss_{\text{FAM}}$(Equation~\eqref{eq:CFAApproximator}) on $\D$ to optimize $g_{\eta}$.
	\State \textbf{Return} $f_{\phi}, g_{\eta}$.
	\end{algorithmic}
\end{algorithm}

The essence of the variance reduction in MBCAL lies in Equation~\eqref{eq:couterfactual_future_advantage}, where the subtraction eliminates the noises from user feedbacks and other sources. We borrow ideas from the advantage function \cite{wang2015dueling}, however, CFA is different from the advantage function in that we do not resample the trajectory but we keep the rest of trajectory (that is $\vo\nolimits_{[t+1:T]}$) unchanged. Although this could bring severe biases in many MDP problems, we argue that the recommender systems embrace weaker correlations between sequential decisions than the other problems (such as robot control and game control). Additionally, as the FAM averages out CFA across different trajectories, the bias turn out to be negligible compared with the benefits of reducing the variances.

\section{Experiments}

\subsection{Datasets}
Evaluation of RL-based recommender systems is challenging. The most convincing metric requires running online A/B tests, but it is not only too costly but also too risky to compare all the baselines in an online system. Offline evaluation of long-term utility using user logs is tricky as we can not have the feedback if the agent recommends a different item than what was stored in the log. In order to thoroughly study the performance of the proposed systems, we follow the previous works to build simulators \cite{dulac2015deep, zhao2018deep, liu2019personalized, Eugene2019RecSim}. But instead of synthetic simulators, we use real-data-driven simulators. 
%Many previous works on RL-based RS use off-line evaluation criteria, including NDCG and Precision \cite{zhao2018deep, liu2018deep} to evaluate the performance. However, such criteria are incapable of indicating the ability of RS to pursue long term rewards. Others build synthetic simulators or real-data driven simulators \cite{dulac2015deep, zhao2018deep, liu2019personalized, Eugene2019RecSim}. In order to thoroughly study the performance of the proposed systems, we use real data driven simulators for Evaluation. 
The datasets used include: MovieLens ml-20m\footnote{\url{http://files.grouplens.org/datasets/movielens/ml-20m-README.html}} \cite{harper2016movielens}, Netflix Prize\footnote{\url{https://www.kaggle.com/netflix-inc/netflix-prize-data}} and NewsFeed\footnote{Data collected from Baidu App News Feed System}, as shown in Table~\ref{tab:dataset}. Details of the datasets are explained as follows.
\begin{itemize}
\item{\emph{MovieLens ml-20m}}:
The dataset describes 5-star rating activities from MovieLens. The user behavior $\Beh = [0,1,2,3,4,5]$ corresponds to the star ratings, with the reward to be $\RFunc = [0,1,2,3,4,5]$. There are 3 kinds of features (\emph{movie-id}, \emph{movie-genre} and \emph{movie-tag}).
\item{\emph{Netflix Prize}}: 
The dataset is a 5-star rating dataset from Netflix. The reward follow the setting of MovieLens. There are only 1 type of features (\emph{movie-id}).
\item{\emph{NewsFeed}}: 
The dataset is collected from a real online news recommendation system. We focus on predicting the dwelling time on the clicked news. The dwelling time is partitioned into 12 levels (e.g., dwelling time < 30, 30 < dwelling time < 40, ...), corresponding to 12 different user behaviors, with the corresponding rewards to be $\RFunc = [1,2,...,12]$. There are 7 kinds of features (\emph{news-id}, \emph{news-tag}, \emph{news-title}, \emph{news-category}, \emph{news-topics}, \emph{news-type}, \emph{news-source}).
\end{itemize}
%\begin{table*}[h!]
%\begin{center} 
%  \caption{Properties of the used datasets.}
%  \begin{tabular}{c|c c c c c c}
%    \hline
%    \textbf{Dataset} & \# of Users & \# of Items & \tabincell{c}{\# of Training \\ Records} & \tabincell{c}{\# of Validation \\ Records} & Item Features & User Behaviors($\Beh$) \\
%    \hline \hline
%    % MovieLens(ml-20m) & 138,493 & 27,278 & 2,486,840 & 1,270,879 & movie-id, genres, tags & ratings(5 classes) \\
%    MovieLens(ml-20m) & 130,000+ & 20,000+ & 2,486,840 & 1,270,879 & movie-id, genres, tags & ratings(5 classes) \\
%    \hline
%    Netflix & 480,000+ & 17,000+ & 4,533,213 & 2,275,125 & movie-id & ratings(5 classes) \\
%    \hline
%    NewsFeed & 920,000+ & 110,000+ & 9,408,891 & 4,700,894 & \tabincell{c}{news-id, tags, title\\category, topics\\news-type, news-source} & dwelling time(12 classes) \\
%    \hline \hline
%  \end{tabular}
%  \label{tab:dataset}
%\end{center}
%\end{table*}

\subsection{Experimental Settings}

\subsubsection{Simulator Details} 
In order to fairly evaluate different methods, it is necessary to avoid the agent in the evaluated system to "hack" the simulator. For this purpose, we add two special settings in the evaluation processes. First, all the agents to be evaluated are allowed to use only a \emph{subset} of features, while the simulator uses the full feature set. In \emph{MovieLens} and \emph{Netflix}, only 1 feature (\emph{movie-id}) is used in the agents. In \emph{NewsFeed}, 4 kinds out of 7 are used (\emph{news-id}, \emph{category}, \emph{news-type}, and \emph{news-source}). Second, we artificially set the model architecture of the simulator to be different from that of the agents. We use the LSTM unit for the simulators, while GRU is used in the agents. To get a view of how close the simulator is to the real environment, we list the micro-F1, weighted-F1, and RMSE with respect to the accuracy of user behavior classification. Properties of datasets and simulators are shown in Table~\ref{tab:dataset}.\footnote{The source code can be found at: https://github.com/LihangLiu/MBCAL} In \emph{NewsFeed}, we also retrieved over $400$ historical A-B test records online. Concerning long-term rewards, including total clicks or dwelling time of a session, the correlation of prediction of our simulators to the real case is $0.90+$.

\begin{table}[h!]
\begin{center} 
  \caption{Properties of Datasets and Simulators.}
  \begin{tabular}{c|c c c}
    \hline
    \textbf{Properties} & \textbf{\emph{MovieLens}} & \textbf{\emph{Netflix}} & \textbf{\emph{NewsFeed}} \\
    \hline
    \textbf{\# of Users} & 130K & 480K & 920K\\
    \textbf{\# of Items} & 20K & 17K & 110K \\
    \textbf{\# of Different Labels} & 6 & 6 & 12\\
    \textbf{\# of Types of Features} & 3 & 1 & 7 \\
    \textbf{Size of Training Set} & 2.48M & 4.53M & 9.41M \\
    \textbf{Size of Validation Set} & 1.27M & 2.27M & 4.70M \\
%    \# of Behavior classes & 5-level ratings & 5-level ratings & 12-level dwelling time\\
%    Feature Set & \tabincell{c}{movie-id, genres,\\ tags} & movie-id & \tabincell{c}{news-id, tags, title, category, topics,\\ news-type, news-source} \\
    \hline
%    \textbf{Simulator Properties}\\
%    \hline
    \textbf{Simulator Macro-F1} & 0.545 & 0.511 & 0.923\\
    \textbf{Simulator Weighted-F1} & 0.532 & 0.498 & 0.887\\
    \textbf{Simulator RMSE} & 0.770 & 0.848 & 1.810\\
    \hline
  \end{tabular}
  \label{tab:dataset}
\end{center}
\end{table}

\subsubsection{Evaluation Settings}
There are two types of iterations in the evaluation: the training round and the test round. For a training round, the agent to be evaluated produces actions by using $\epsilon$-greedy policy ($\epsilon=0.1$ throughout all experiments). It then updates its policy using the collected feedback from the simulator. In the test round, the algorithm produces actions by using a greedy policy, which is evaluated by the simulator. The data generated in the test round are not used in training. For each session in training or test rounds, we consider $T=20$ steps of interaction between the simulator and the agent. Each training round or test round includes 256,000 sessions.

\begin{figure}[t!]
\centering
\includegraphics[width=0.95\columnwidth]{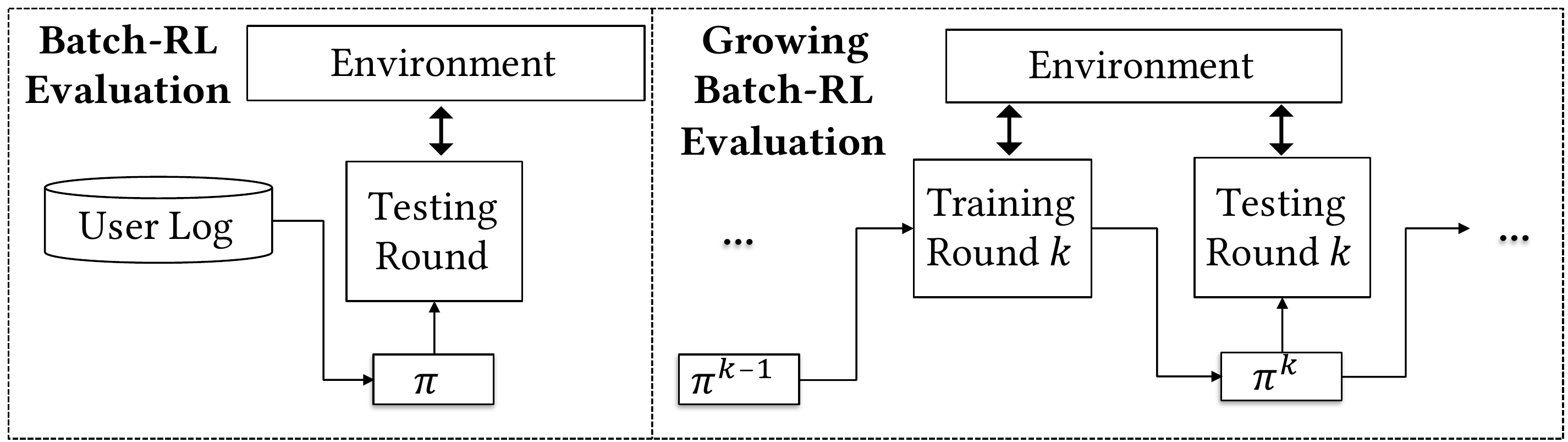}
\caption{The evaluation processes.}
\label{fig:Evaluation}
\end{figure}

For each experiment, we report the \emph{average reward per session} in the test round, defined by $1/|\D_{test}| \cdot \sum_{\vo \in \D_{test}}\sum_{t=1}^T r_t$, where $\D_{test}$ is the collection of trajectories in test round.
Each experiment is repeated with different random seed for three times, the mean and variance of the score is reported. We also simulate the \emph{Batch RL} and the \emph{Growing Batch-RL} evaluation separately (Figure~\ref{fig:Evaluation}). In the \emph{Batch RL} evaluation, the agent can use only the static user log for training, and it interacts with the simulator for test. In \emph{Growing Batch RL} evaluation, for both training round and test round, the agent needs to interact with the simulator. The training round repeats up to 40 times.

\subsection{Methods for Comparison}

We compare different methods ranging from Supervised Learning (GRU4Rec), bandits (GRU4Rec ($\epsilon$-greedy)) to MFRL (MCPE, DQN, DDQN, and DDPG) and MBRL (Dyna-Q). For bandits, LinUCB \cite{Chu2011Contextual} is commonly used as a baseline. However, in our environments, LinUCB performs poorly due to the insufficient representative power of the linear model. Thus, we post the results of $\epsilon$-greedy version of NN models (GRU4Rec ($\epsilon$-greedy)) instead of LinUCB.

The methods for comparison include:
\begin{itemize}
    \item \textbf{GRU4Rec} \cite{Hidasi2015Session}: It adopts GRU to encode the interactive history to predict the instant user behavior. The model architecture is equivalent to the environment model. We use the entropy losses in GRU4Rec.
    \item \textbf{GRU4Rec ($\epsilon$-greedy)}: It applies the $\epsilon$-greedy selection of items in GRU4Rec during the training rounds.
    \item \textbf{DQN} \cite{zheng2018drn}: A classical off-policy learning algorithm \cite{mnih2015human}. For state representation, to ensure a fair comparison between different learning algorithms, GRU is used to encode the historical observations, which is equivalent to GRU4Rec and our method.
    \item \textbf{DDQN} \cite{zheng2018drn}: Double DQN \cite{Hasselt2016DDQN} uses a different action selection for value backup to avoid the value overestimation in off-policy learning. The model architecture is kept equivalent to GRU4Rec.
    \item \textbf{DDPG} \cite{zhao2018deep}: Deep Deterministic Policy Gradient (DDPG) \cite{lillicrap2015continuous} is an off-policy learning algorithm for continuous action space. The inferred action is used to select the item that lies closest to it for display (nearest neighbor). We use the same neural structure as GRU4Rec for both the actor and critic networks in DDPG.
    \item \textbf{MCPE}: Monte Carlo Policy Evaluation \cite{Dimitrakakis2008Roll} is a straight-forward value iteration algorithm. The Monte Carlo evaluation of the whole trajectory is used, i.e., we apply Equation~\eqref{eq:ValueIteration} with $\hat{Q}_{\text{target}} = \sum_t r_t$. Again, we keep the model architecture equal to the other baselines.
    \item \textbf{Dyna-Q} \cite{liu2019personalized, zoupseudo}: Dyna-Q is an MBRL method that augments DQN with the imagined rollouts from an environment model. The ratio of imagined rollouts to real trajectories is set to 1:1.
    \item \textbf{MBCAL}: The full version of our method.
    \item \textbf{MBCAL (w/o variance reduction)}: It is an ablative version of MBCAL. We use SFR instead of CFA as the label of FAM, i.e., in Equation~\eqref{eq:CFAApproximator}, $\hat{R}^{\text{future}}_{\phi}$ is used instead of $\hat{A}^{\text{future}}_{\phi}$.
\end{itemize}

All the parameters are optimized by adam optimizer with learning rate = $10^{-3}$, $\beta_1 = 0.9$ and $\beta_2 = 0.999$. The decay factor in the long-term reward is set to be $\gamma=0.95$. The embedding sizes for the \emph{item-id} and other id type features are all set to 32. The hidden size for MLP is set to 32. For training MEM in MBCAL, we use $p_{mask}=0.20$ to generate the masked trajectories $D_M$. In DDPG, we found that high dimensional action space yields really poor performance. Thus we use a 4-dimensional action space. Correspondingly we use an additional layer to map the item representation into a 4-dimensional vector.

\subsection{Experimental Results}

\begin{table}[h!]
\begin{center} 
  \caption{Average reward per session of different algorithms and datasets in Batch-RL evaluation.}
  \begin{tabular}{c|ccc}
    \hline
    \multirow{2}{*}{\textbf{Algorithms}} & \multicolumn{3}{c}{\textbf{Average reward per session}}\\
    \cline{2-4}
    & \textbf{\emph{Movielens}} & \textbf{\emph{Netflix}} & \textbf{\emph{NewsFeed}} \\
    \hline
    \textbf{GRU4Rec}     & $77.93\pm0.06$ & $79.63\pm0.02$ & $11.58\pm0.14$ \\
    \textbf{DDPG}        & $70.99\pm0.70$ & $72.50\pm0.35$ & $10.90\pm0.42$ \\
    \textbf{DQN}         & $77.27\pm0.06$ & $77.75\pm0.01$ & $12.44\pm0.33$ \\
    \textbf{DDQN}        & $77.23\pm0.02$ & $77.70\pm0.04$ & $12.48\pm0.17$ \\
    \textbf{MCPE}        & $77.20\pm0.10$ & $77.70\pm0.03$ & $13.21\pm0.53$ \\
    \textbf{Dyna-Q}      & $77.25\pm0.05$ & $77.81\pm0.02$ & $13.04\pm0.33$ \\
    \textbf{MBCAL}       & $\textbf{78.02}\pm0.03$ & $\textbf{79.71}\pm0.04$ & $\textbf{16.32}\pm0.24$ \\
    \tabincell{c}{\textbf{MBCAL}(w/o\\ variance reduction)} & $77.70\pm0.04$ & $79.50\pm0.04$ & $15.61\pm0.38$ \\
    \hline
  \end{tabular}
  \label{tab:simB}
\end{center}
\end{table}

\subsubsection{Results of Batch-RL Evaluation} The results on Batch-RL evaluation are shown in Tab.~\ref{tab:simB}. We evaluate the reward of a session according to the reward generated by the simulator. To conclude from the result, MFRL can not compete with MBRL in all three environments. As MFRL is sample inefficient, it tends to have poor startup performance. Surprisingly DDPG has the weakest performance in all three environments. By carefully investigating the value functions in DDPG, we found that DDPG overestimates the value function a lot compared with the other MFRL. We thought that the overestimation comes from value backups from continuous actions that may not correspond to realistic items. The overestimation problem in actor-critic methods has also been thoroughly investigated \cite{Fujimoto18Addressing}.

As is expected, the MBCAL leads the performance of all the tested systems with substantial margins, demonstrating its sample efficiency. However, for \emph{Movielens} and \emph{Netflix}, our method earns a smaller margin over the supervised learning method compared with that of \emph{NewsFeed}. It is likely that the long term reward plays a more significant role in \emph{NewsFeed} than the other two environments. Furthermore, as learning to predict long-term utility requires more data than the instant reward, the preponderance of RL has not yet been sufficiently revealed in Batch-RL settings. However, it is essential that the performance of MBCAL at the start stage is already state-of-the-art, which proves that MBCAL has low risks and high sample efficiency.

\begin{figure*}
\centering
    \begin{subfigure}{.666\columnwidth}
        \includegraphics[width=1.0\columnwidth]{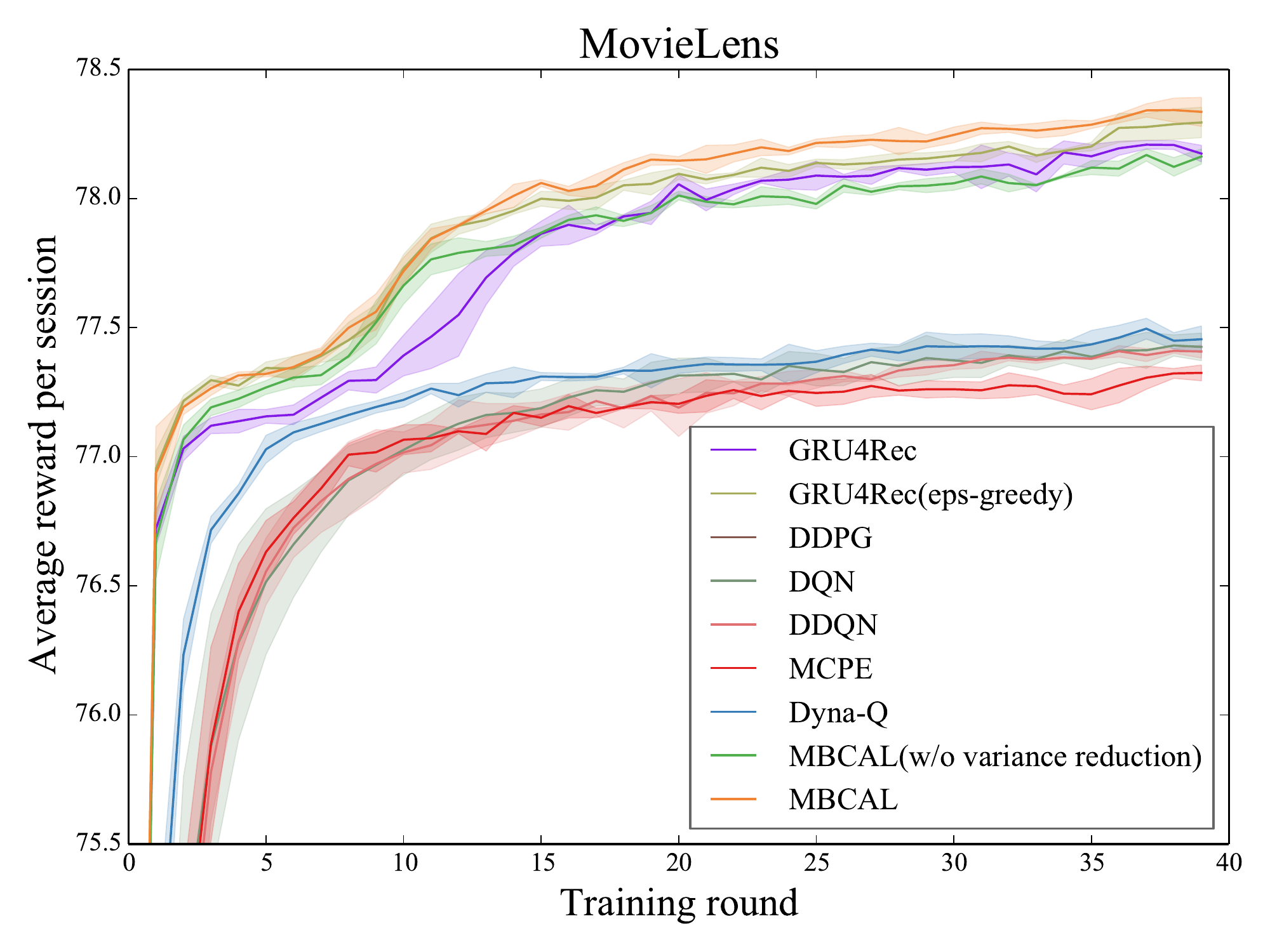}
        % \caption{Score of different mask steps on Movielens.}
        \label{fig:exp_res:a}
    \end{subfigure}\hfill
    \begin{subfigure}{.666\columnwidth}
        \includegraphics[width=1.0\columnwidth]{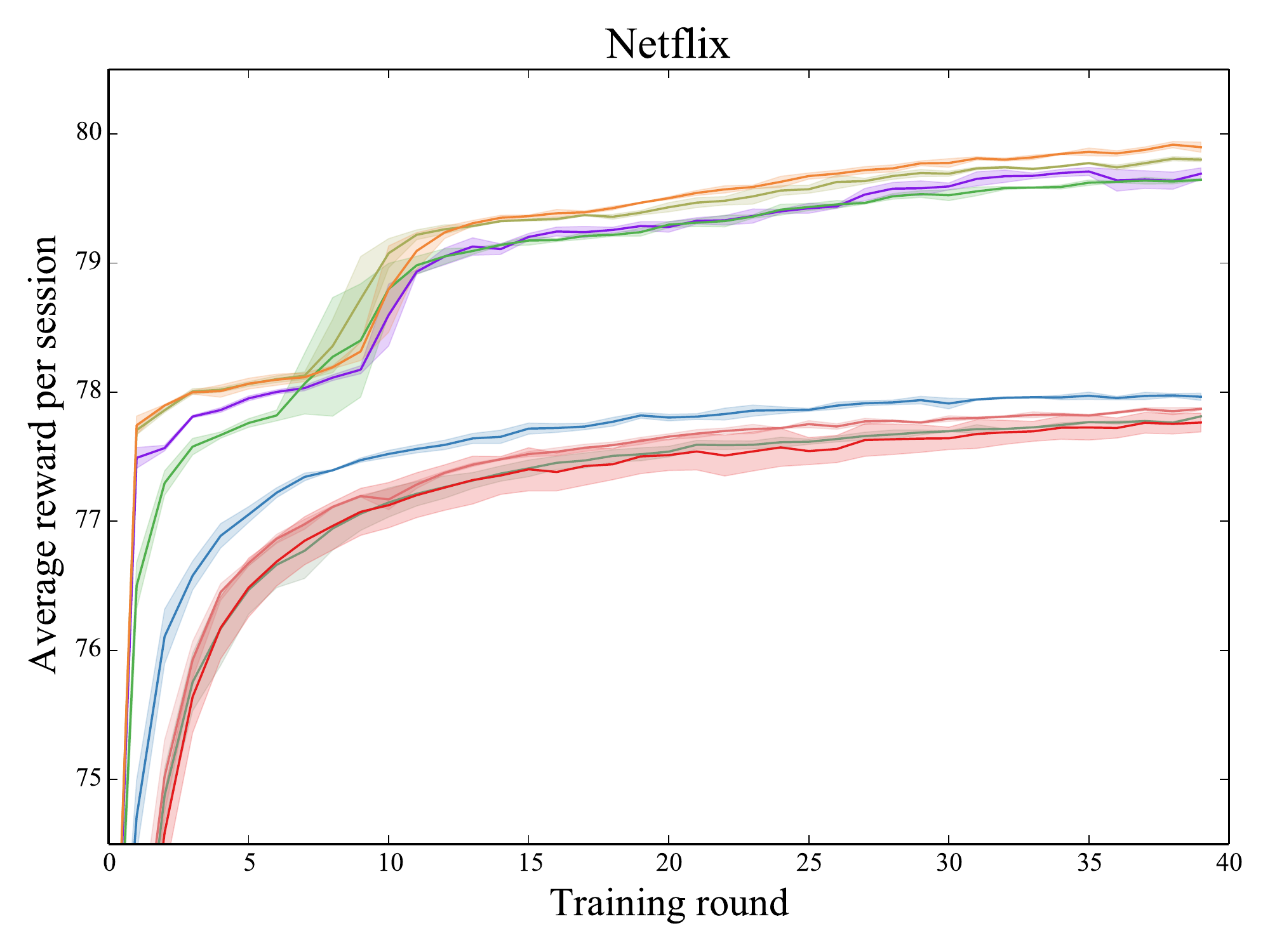}
        % \caption{Score of different mask steps on Netflix.}
        \label{fig:exp_res:b}
    \end{subfigure}\hfill
    \begin{subfigure}{.666\columnwidth}
        \includegraphics[width=1.0\columnwidth]{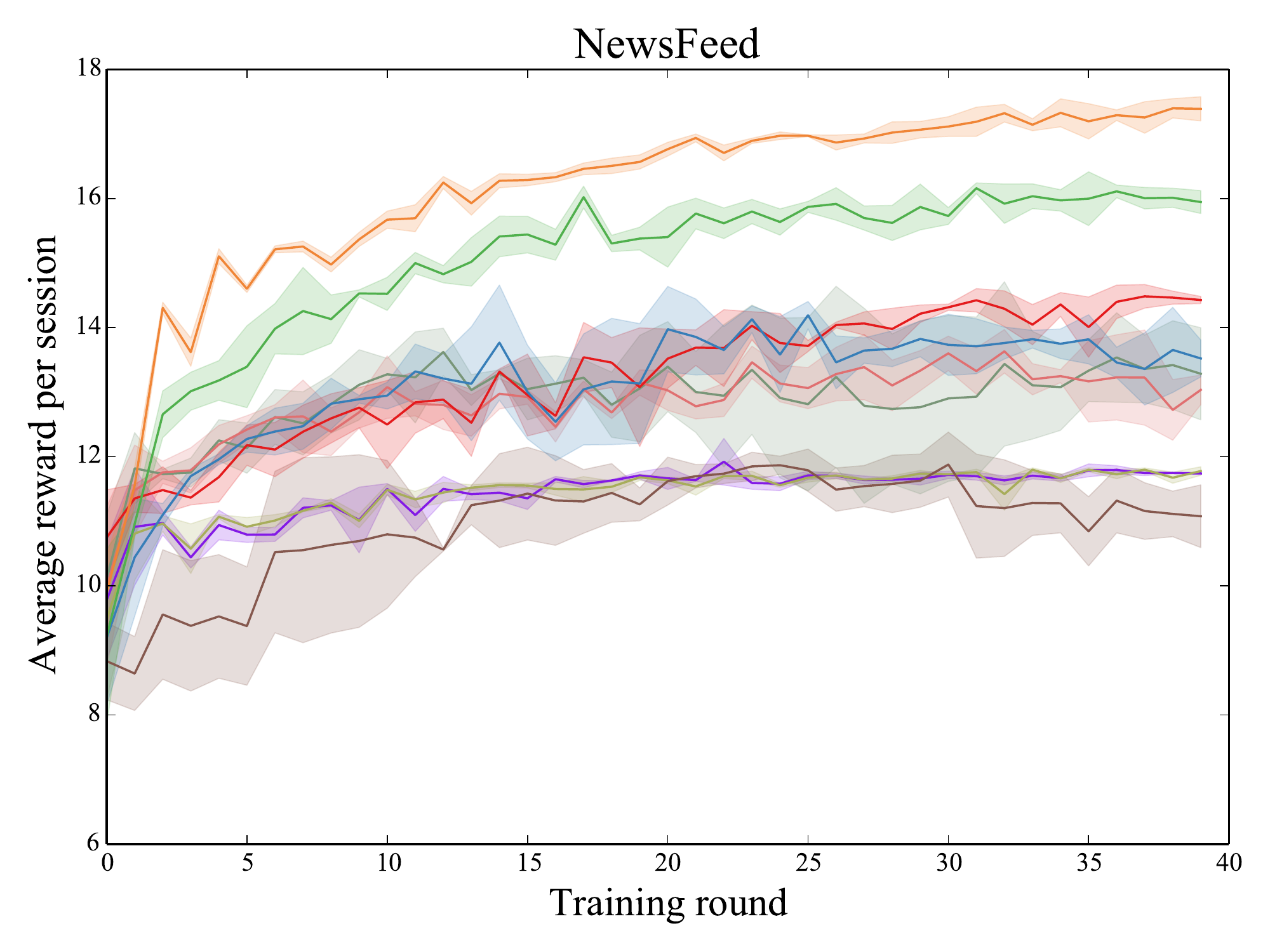}
        % \caption{Score of different mask steps on FeedGR.}
        \label{fig:exp_res:c}
    \end{subfigure}
\caption{Average reward per session of different algorithms and datasets in Growing Batch-RL evaluation, with the horizontal axis represent the training rounds.}
\label{fig:exp_res}
\end{figure*}

\subsubsection{Results of Growing Batch-RL Evaluation} Figure~\ref{fig:exp_res} shows the results of the Online Evaluation in three environments. GRU4Rec($\epsilon$-greedy) surpasses the purely supervised learning GRU4Rec by a small margin in every environment, showing the benefit of exploration in online systems. Performances of DDPG in all three environments are again surprisingly bad. In Figure~\ref{fig:exp_res}, we show DDPG curve in \emph{NewsFeed} environment only because in the other two environments, DDPG lags too much behind all the other methods. We believe that continuous action space for recommendation systems with dynamic discrete item space can not work well enough.

With the assistance of the environment model, Dyna-Q gains some advantages at the beginning, but it gradually subsides as the learning continues. This phenomenon is just in line with the expectations, for the virtual experience quickly loses its benefit with the accumulation of sufficient real user feedback. MBCAL again keeps its performance ahead of the other methods in all the environments. Even for \emph{Netflix} and \emph{Movielens}, where the other RL-based system fails to gain any benefits over traditional GRU4Rec, MBCAL wins with a considerable margin. In \emph{NewsFeed}, where the long term rewards play a more critical role, MBCAL strengthens the leading edge.

MCPE, DQN, DDQN, and Dyna-Q lag entirely behind the other methods, including supervised learning baselines in \emph{Movielens} and \emph{Netflix} environment, while this is not true in \emph{NewsFeed}. We investigate the reason by setting the output of GRU4Rec to the instant reward instead of the user behavior classification, which turned the classification into regression, and the entropy loss to mean square error loss. We found a significant drop in performance in GRU4Rec, which is more consistent with the results in \emph{NewsFeed}. The results show that classification and entropy loss benefit the system more than regressions.  An explanation is that user behavior contains more abundant information than the rewards, which also made MBRL more advantageous than MFRL.
%However, most RL-based, especially value iteration algorithms approximate the rewards only. It turns out that such settings can lower the performance significantly in some recommender systems. 

\subsubsection{Analysis of the variance} The key point in MBCAL is the variance reduction through counterfactual comparisons. The previous proposals \cite{Geman92Neural} suggest that the mean square error (MSE) in a well-trained model is composed of the model bias and the variance(noise) in the labels. As we use equivalent neural architectures in all the methods for comparison, they share the same model bias. Therefore the mean square error shall be dominated by the noise. To study whether CFA truly reduces the variance, we compare the MSE from Equation~\eqref{eq:ValueIteration} and Equation~\eqref{eq:CFAApproximator}. We compare the MSE of MCPE, DQN, Dyna-Q, MBCAL (w/o variance reduction), and MBCAL, based on the interactive logs collected in the test round of Batch-RL evaluation. The average MSE is presented in Table~\ref{tab:train_variance}.

\begin{table}[h!]
\begin{center} 
  \caption{The mean square error (MSE) loss of different algorithms in different environments.}
  \begin{tabular}{c|ccc}
    \hline
    \multirow{2}{*}{\textbf{Algorithms}} & \multicolumn{3}{c}{\textbf{MSE loss}}\\
    \cline{2-4}
     & \textbf{\emph{Movielens}} & \textbf{\emph{Netflix}} & \textbf{\emph{NewsFeed}} \\
    \hline
    \textbf{DQN}         & 1.50 & 1.22  & 4.29  \\
    \textbf{MCPE}        & 17.1 & 9.21 & 46.9 \\
    \textbf{Dyna-Q}      & 0.94 & 1.04 & 7.87 \\
    \textbf{MBCAL}       & \textbf{0.004} & \textbf{0.009} & \textbf{0.07} \\
    \tabincell{c}{\textbf{MBCAL} (w/o\\ variance reduction)} & 3.45 & 3.29  & 3.07 \\
    \hline
  \end{tabular}
  \label{tab:train_variance}
\end{center}
\end{table}

According to the previous theoretical analysis, using value backup of longer horizons suffer from more significant variance. The variance of MCPE is indeed higher than that of DQN and Dyna-Q, as backup of the whole trajectory is used. The MBCAL (w/o variance reduction) has the second-largest variance. It is smaller compared with MCPE because using simulated rollout from the environment model already eliminates part of the noises. The variances of DQN and Dyna-Q are smaller because one-step value backup is employed. Compared with the other methods, MBCAL embraces significantly lower variances, which shows that variance has been reduced as expected.

\section{Conclusion}

To conclude this work, we are focused on the sequential decision-making problems in the recommender systems. To maximize the long-term utility, we propose a sample efficient and variance reduced reinforcement learning method: MBCAL. It involves a masked environment model to capture the instant user behavior, and a future advantage model to capture the future utility. Through counterfactual comparison, MBCAL significantly reduces the variance in learning. Experiments on real-data-driven simulations show that the proposed method transcends the previous ones in both sample efficiency and asymptotic performances. Possible future extensions to this work may be to theoretically calculating the error bound, and to extend the fixed horizon settings to infinite and dynamic horizon recommender systems.

\bibliographystyle{ACM-Reference-Format}
\bibliography{references}
\clearpage

\end{document}